\documentclass[amsmath,amssymb,aps,prb]{revtex4-2}
\usepackage{braket}
\usepackage{graphicx}

\begin{document}

\title{Field and temperature tuning of magnetic diode in permalloy honeycomb lattice}

\author{George Yumnam$^1$}
\author{Moudip Nandi$^1$}
\author{Pousali Ghosh$^1$}
\author{Amjed Abdullah$^2$}
\author{Mahmoud Almasri$^2$}
\author{Erik Henriksen$^3$}
\author{Deepak K. Singh$^{1}$}
\email{singhdk@missouri.edu}
\affiliation{$^1$ Department of Physics and Astronomy, University of Missouri, Columbia, Missouri 65211, USA}
\affiliation{$^2$ Department of Electrical Engineering and Computer Science, University of Missouri, Columbia, Missouri 65211, USA}
\affiliation{$^3$ Department of Physics, Washington University in St. Louis, Missouri, 63130, USA}

\begin{abstract}
The observation of magnetic diode behavior with ultra-low forward voltage of 5 mV renders new venue for energetically efficient spintronic device research in the unconventional system of two-dimensional permalloy honeycomb lattice. Detailed understanding of temperature and magnetic field tuning of diode behavior is imperative to any practical application. Here, we report a comprehensive study in this regard by performing electrical measurements on magnetic diode sample as functions of temperature and magnetic field. Magnetic diode is found to persist across the broad temperature range. Magnetic field application unveils a peculiar reentrant characteristic where diode behavior is suppressed in remnant field but reappears after warming to room temperature. Analysis of $I-V$ data suggests a modest energy gap, $\sim$0.03 - 0.1 eV, which is comparable to magnetic Coulomb’s interaction energy between emergent magnetic charges on honeycomb vertices in the reverse biased state. It affirms the role of magnetic charge correlation in unidirectional conduction in 2D honeycomb lattice. The experimental results are expected to pave way for the utilization of magnetic diode in next generation spintronic device applications.
\end{abstract}

\maketitle

\section{Introduction}
Spin or magnetic diode is a critical component to the spintronic device design for information processing and logic operations.\cite{ref1,ref2,ref3,ref4} Therefore, a lot of emphasis is made to develop a practical prototype of magnetic diode, which could function at room temperature. Previously, researchers have used the concept of spin-charge interaction mechanism to develop a magnetic diode.\cite{ref5,ref6,ref7,ref8} Several magnetic materials and systems are intensively investigated in the past that can be artificially tailored to depict the asymmetric spin population. It includes multilayered magnetic devices that exhibit a very high level of unidirectional spin polarization and the design of bipolar and unipolar spin diodes across a magnetic domain wall or magnetic tunnel junction.\cite{ref9,ref10,ref11,ref12,ref13,ref14} However, application of magnetic field is universally required. For the first time, we demonstrated the magnetic diode effect at room temperature in an unconventional system of two-dimensional geometrically frustrated permalloy honeycomb lattice that can function without magnetic field application.\cite{ref15,ref16}

The honeycomb specimen, made of ultra-small permalloy elements with typical size of 11 nm$\times$4 nm$\times$8.5 nm, manifests thermally tunable characteristic due to the moderate inter-elemental magnetic dipolar interaction $\sim$ 30 – 40 K.\cite{ref17,ref18} Information about the temperature and magnetic field tuning properties of magnetic diode are highly desirable for any practical application, but are not known yet. In this article, we report comprehensive investigation of electrical properties of magnetic diode using four-probe measurements at different temperatures and magnetic field. Detailed study unveils several new properties that not only sets it apart from a conventional semiconductor diode,\cite{ref19} but also sheds new light on the underlying weak semiconducting characteristic of the magnetic lattice. It includes the demonstration of ultra-small forward threshold voltage (mV range) with diminishing magnitude as a function of temperature, an unusual freezing of magnetic charges in remnant state and an unequivocal connection between the electrical gap energy (mathematically equivalent to the Arrhenius energy) and the magnetic Coulomb interaction between emergent magnetic charges in the reverse biased state.

An artificial magnetic honeycomb is an archetypal two-dimensional geometrically frustrated magnet. In 2D permalloy (Ni$_{0.81}$Fe$_{0.19}$) honeycomb, magnetic moments align along the length of constituting element due to large shape anisotropy.\cite{ref20} Consequently, two-types of local moment arrangements of two-in \& one-out (or vice-versa) or, all-in or all-out states, arise on honeycomb vertices.\cite{ref21,ref22}  Under the Dumbbell formalism, a magnetic moment is considered to be made up of a pair of ‘+q’ and ‘-q' charges, also called magnetic charges, that interact via magnetic Coulomb’s interaction.\cite{ref23} Here q is given by $M/L$, $M$ being the net magnetic moment along the length $L$ of honeycomb element. The algebraic summation of magnetic charges on a given vertex results in the two integer multiplicities of $\mid$Q$\mid$ and $\mid$3Q$\mid$ units for the local magnetic configurations of two-in \& one-out (or vice-versa) or, all-in or all-out, respectively.\cite{ref17} 3Q charges are highly energetic, thus unstable.\cite{ref24} In general, the magnetic charges are highly mobile in thermally tunable environment.\cite{ref25} The charge pattern across the lattice is constantly seeking a low entropic configuration by emitting or absorbing the dynamic magnetic charge defect Q$_m$ of magnitude $\mid$2Q$\mid$ unit, which is the difference between energetic $\mid$3Q$\mid$ charge and the low energy $\mid$Q$\mid$ charge.\cite{ref20,ref24,ref26} 

\begin{figure*}
\centering
\includegraphics[width=\textwidth]{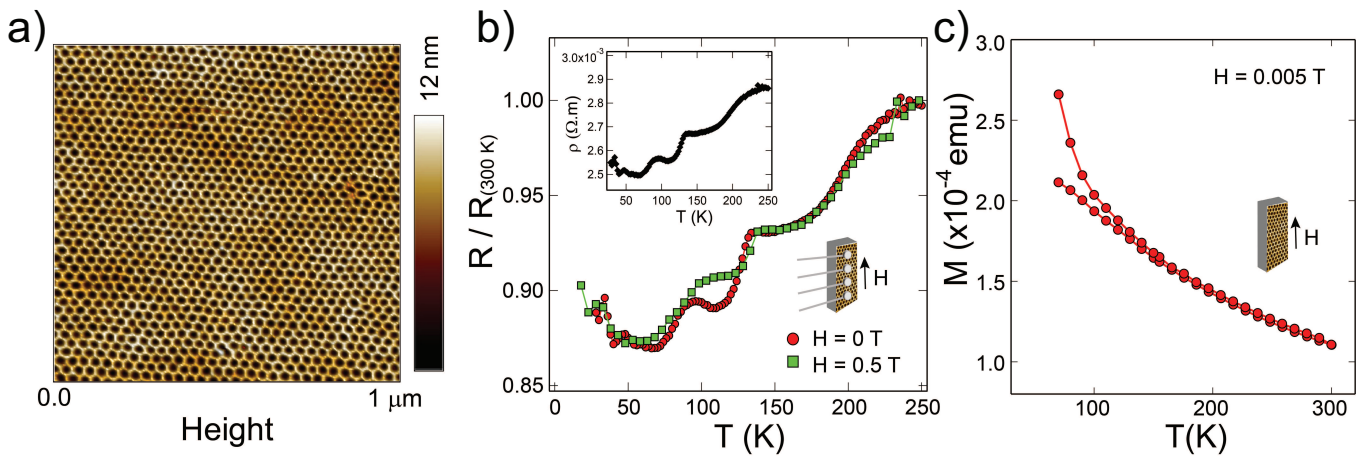} 
\caption{(color online) Topographical image of magnetic diode honeycomb sample and electrical properties. (a) Atomic force micrograph of artificial honeycomb lattice showing high structural fidelity. (b) Resistance vs. temperature plot of $\sim$8.5 nm thick permalloy honeycomb lattice in zero and applied field $H$ = 0.5 T. Magnetic field is applied in-plane to the substrate. The field magnitude is larger than the coercivity of the sample. (Inset shows the plot of resistivity ($\rho$) as a function of temperature in zero field). (c) Magnetization measurement at $H$ = 50 Oe, applied in-plane to the sample, reveals paramagnetic characteristic of permalloy honeycomb lattice. 
} 
\end{figure*}

In recent studies, it was shown that the charge defect’s dynamics or magnetic charge fluctuation on honeycomb vertices generate transverse fluctuations in local magnetic field $\mathbf{B(k)}$ that couple to conduction electron’s spin $\sigma$.\cite{ref27} The interaction is found to spur magnetic charge propelled electrical conduction in nanoscopic permalloy honeycomb lattice.\cite{ref28} Besides the propelled conduction, the thermally tunable permalloy honeycomb also exhibits magnetic diode behavior, characterized by the unidirectional conduction at ultra-small threshold voltage.\cite{ref15} A detailed investigation of the underlying mechanism behind the unidirectional conduction revealed the symmetry defying property of honeycomb lattice in the current biased states.\cite{ref16} The lattice vertices are found to be mainly occupied by the low (high) multiplicity $\pm$Q ($\pm$3Q) charges in the forward (reverse) biased states. Theoretical calculations under the reasonable assumption of drift diffusion formalism\cite{ref29} unequivocally showed that while the low multiplicity charges facilitate conduction, the high multiplicity $\pm$3Q charges inhibit electrical transport process by creating an energy barrier.\cite{ref30} Consequently, a diode-type electrical property emerges in the thermally tunable permalloy honeycomb lattice with direct linkage to the magnetic charge physics. Here, we focus on the important practical aspects of magnetic diode e.g. temperature and magnetic field tuning that can pave way for practical applications in the spintronic device designs.

\section{Results and Discussion}

The fabrication process of large throughput nanoscopic honeycomb lattice, as described in detail in the experimental section, involves multiple steps that include the development of nanoporous polymer template on top of a silicon substrate,\cite{ref31} followed by the reactive ion etching to transfer the template to the underlying silicon substrate and permalloy material deposition on the patterned substrate in near parallel configuration. In Fig. 1a, we show the atomic force micrograph of a typical artificial honeycomb lattice resulting from the nanofabrication process, which confirms the high quality of the sample. The large specimen size allows for the bulk properties investigation using various macroscopic probes, such as electrical and magnetic measurements. In Fig. 1b-c, we show the plots of electrical resistance and magnetization as a function of temperature. While magnetic data exhibits paramagnetic characteristic, electrical measurements manifest several peculiarities. The paramagnetic behavior is also consistent with the previous neutron scattering-based study, which revealed a massively degenerate ground state of magnetic charges in the system.\cite{ref32} In electrical measurements, first we notice that the resistivity at room temperature is high. Such high resistivity is typically found in semiconductors.\cite{ref19} Second, normalized resistance (R(T)/R(300 K)) depicts complex temperature dependence in both zero and applied magnetic field. As temperature decreases, resistance decreases and develops a plateau around $T$ = 150 K. For further decrease in temperature, the R($T$)/R(300 K) reduces at a faster rate, indicating a phase transition. Similar behavior, albeit more resistive, is also detected in a thinner, $\sim$6 nm, permalloy honeycomb lattice.\cite{ref33} Interestingly, magnetic field application of $H$ = 0.5 T (applied inplane to the sample) diminishes the sharp downturn in resistance, thus altering the electrical characteristic. Finally, the trend is reversed at low temperature by registering a modest increase in resistance as temperature decreases further. 

\begin{figure}
\centering
\includegraphics[width=0.75\textwidth]{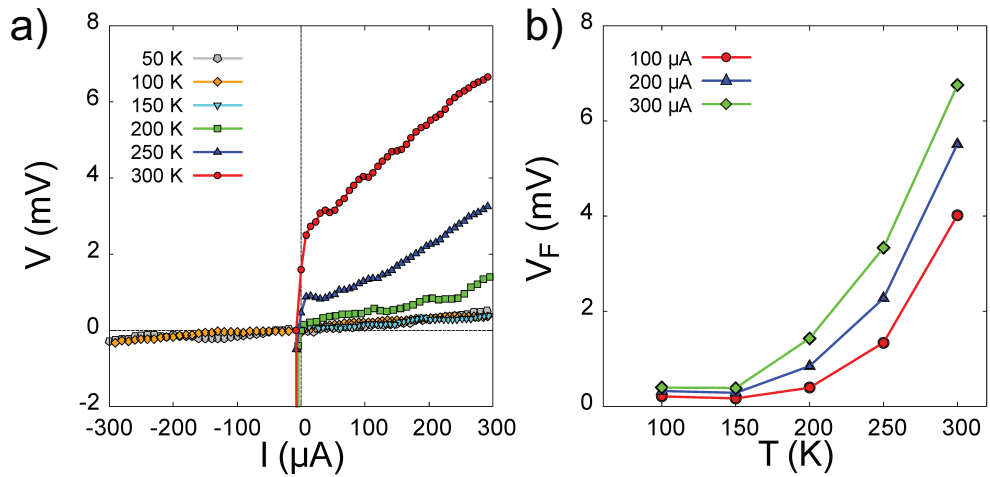} 
\caption{(color online) I-V characteristics of magnetic diode. (a) I-V data obtained in four-probe configuration at few characteristic temperatures. The diode behavior is onset at ultra-low forward threshold voltage of $\sim$mV. Electrical conduction on the reverse side is blocked at I $<$ 10 $\mu$A. (b) Temperature dependence of forward voltage $V_F$ in magnetic diode. $V_F$ is found to decrease as a function of temperature.
} 
\end{figure}

To further understand the electrical properties of permalloy honeycomb lattice, we have performed I-V measurements at different temperatures in zero and applied magnetic field in standard four-probe configuration. Four probe measurement tends to remove the artifact due to contact resistance. We show the I-V plots at different temperatures in Fig. 2a. Several features, resembling diode-type behavior, are immediately noticed. It includes the observation of unidirectional conduction at ultra-low threshold voltage, $\sim$mV and its persistence across a broad temperature range. Interestingly, the forward voltage seems to be reducing as temperature decreases. To quantify this observation, we have plotted the representative forward voltage, V$_{F}$, at several current values as a function of temperature in Fig. 2b. It clearly shows that $V_F$ decreases from 4 mV at $T$ = 300 K to 0.6 mV at $T$ = 150 K. Although the depicted behavior in forward voltage is consistent with the $R$ vs $T$ measurement, shown in Fig. 1c, where electrical resistance is found to be decreasing as a function of temperature, it is in strong contrast to the conventional semiconductor-based diode where the forward voltage increases as a function of temperature.\cite{ref34} It suggests an impending new mechanism behind the unidirectional conduction in permalloy honeycomb lattice.

\begin{figure*}
\centering
\includegraphics[width=\textwidth]{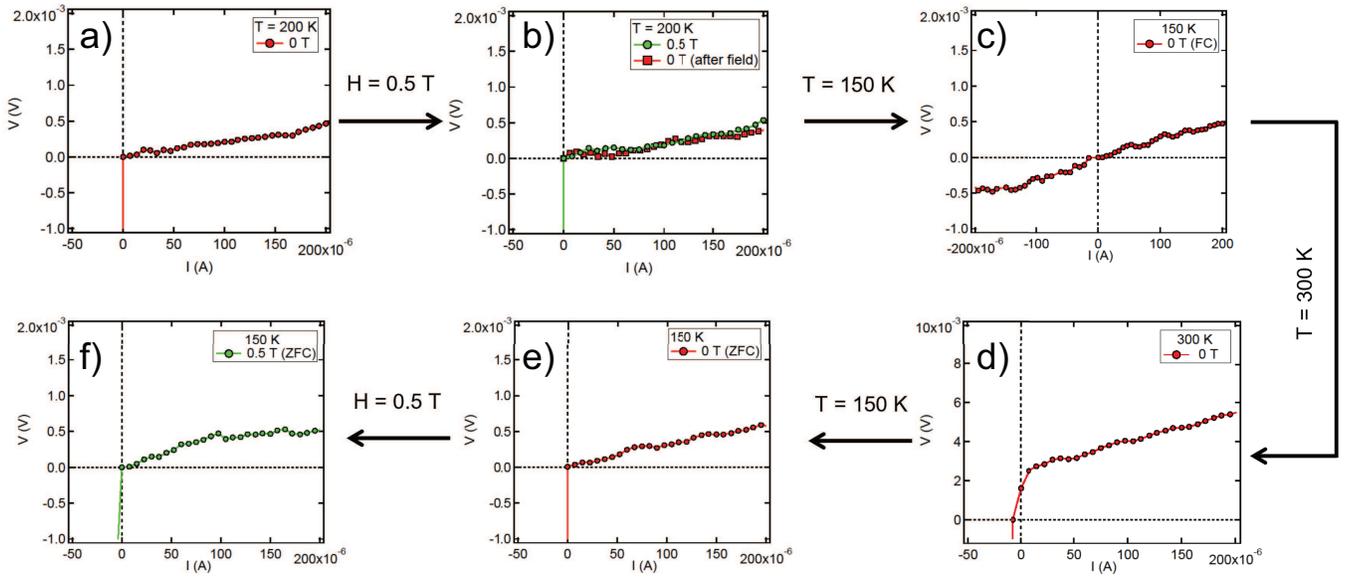} 
\caption{(color online) Magnetic field effect on unidirectional conduction. Comprehensive investigation of field effect reveals electrical memory preservation in permalloy honeycomb lattice. To demonstrate the unusual behavior, we show the I-V measurements in zero, applied and remnant fields at few temperatures. The diode characteristic at $T$ = 200 K (fig. a) remains unperturbed to $H$ = 0.5 T field application (fig. b). Setting the field to $H$ = 0 T does not affect the diode behavior. When the sample is cooled to $T$ = 150 K in the remnant field, the diode behavior disappears (fig. c). However, warming the sample to $T$ = 300 K recovers the diode effect (fig. d). Subsequent cooling of the sample to $T$ = 150 K in zero field, again reveals strong magnetic diode phenomenon (fig. e). Furthermore, setting the field to $H$ = 0.5 T does not change the diode behavior (fig. f) at the same temperature (similar to fig. b).
} 
\end{figure*}

The underlying physics in a conventional semiconductor-based diode is dictated by the band gap between valence and conduction bands, which can be tuned by appropriate chemical doping.\cite{ref22} On the other hand, the semiconducting property in magnetic honeycomb lattice is governed by the magnetic charge correlation on honeycomb vertices. Magnetic field application tends to polarize magnetic moment towards the field directions, thus alters the charge pattern by removing the energetic 3Q charges. The implication of field induced magnetic charge rearrangement on unidirectional electrical conduction is not clear. In principle, the absence of 3Q charges on honeycomb vertices would nullify the magnetic diode behavior. However, we observe a more interesting trend in electrical properties in applied magnetic field, see Fig. 3. We find that field application of $H$ = 0.5 T at $T$ = 200 K does not affect the diode behavior (magnetic field was applied along the forward current application direction). But when the field is set to zero and temperature is reduced to $T$ = 150 K in the remnant state, the unidirectional tendency in electrical transport property disappears. Instead of the diode behavior, we observe symmetrical conduction in both the forward and the reverse biased states at $T$ = 150 K. To understand the magnetic field effect, measurements were performed at multiple temperatures in succession: $T$ = 200 K, 150 K, 300 K and back to 150 K. As shown in Fig. 3, magnetic diode behavior is recovered when the sample is warmed to room temperature. Consequent measurements at lower temperature, $T$ = 150 K, in zero field do not seem to affect the diode behavior. Similar observations were made for the field application at $T$ = 250 K and subsequent measurements at lower temperature in the remnant state.

\begin{figure}
\centering
\includegraphics[width=0.75\textwidth]{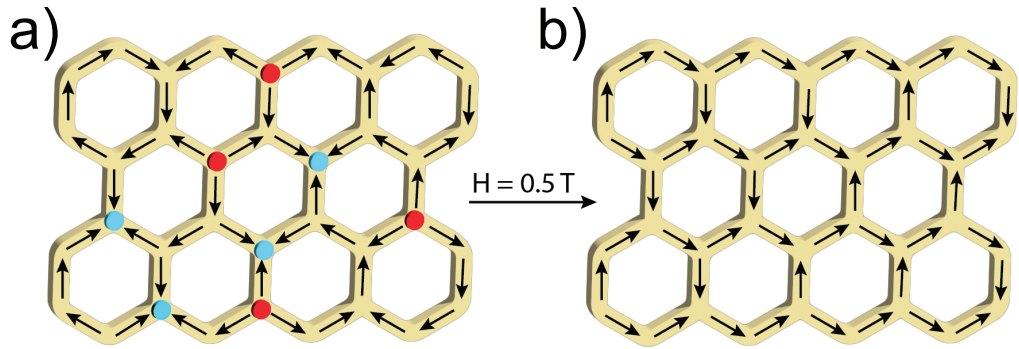} 
\caption{(color online) Schematic of magnetic field application on moment polarization. Magnetic field application aligns magnetic moment towards the field application direction. The magnetic charge configuration in zero field (fig. a) evolves into a new state, comprised of mainly $\pm$Q charges, in applied field (fig. b). Here, we only flip those moment directions that were either aligned opposite to the field application direction (in fig. a) or have a component opposite to it. Moments with perpendicular direction to field are kept unperturbed for the sake of simplicity.
} 
\end{figure}

The unusual field effect on unidirectional conduction in permalloy honeycomb lattice is intriguing. We note that the magnitude of inplane applied magnetic field, $H$ = 0.5 T, is much larger than the coercivity, $H_c$ $\sim$0.1 T, of the sample.\cite{ref16} As reported previously, net magnetization of permalloy honeycomb saturates above $H_c$ $\sim$ 1000 Oe, indicating the strength of magnetic coercivity of the system. Also, it does not seem to be affected by the current application. Thus, the applied field of $H$ = 0.5 T is sufficient to polarize magnetic moment in honeycomb element along the field application direction. Accordingly, the population density of high multiplicity charges on honeycomb vertices will be reduced. It is schematically described in Fig. 4. Yet, the magnetic diode behavior remains unperturbed to magnetic field application. A plausible explanation to this unusual effect, perhaps, lies in understanding the role of thermal fluctuation to moment and magnetic charge fluctuations. The large thermal energy at high temperature overweighs Zeeman’s term, thus inhibits the system from attending the low energy configuration. On the other hand, when sample temperature is reduced, then magnetic moments in remnant state possibly freezes, leading to fewer 3Q charges. This could be a possible reason behind the bi-directional conduction in the remnant state at lower temperature.

Next, we try to understand the mechanism behind the semiconducting characteristic in permalloy honeycomb lattice. This exercise can provide quantitative insight in the energy crossing barrier or gap energy in the reverse biased state. For this purpose, we analyze the $I-V$ traces to extract the energy gap, analogous to the band gap in a conventional semiconductor. The typical diode equation, describing the relation between current and voltage, is given by,
\begin{equation}
I = I_0 \left[\exp(eV/kT) – 1 \right]                 \label{eq1}
\end{equation}

where $I$ is the current through the diode, $I_0$ is the maximum current for a large reverse bias voltage (in this case 24 V), $V$ is the voltage across the diode and $T$ is sample temperature. It is the quantity $I_0$, which is directly dependent on the gap energy $E_g$. Following the research work by Precker et al.,\cite{ref34} we directly write down the formula describing the dependence of $I_0$ on $E_g$ as
\begin{equation}
I_0 = A. T^{(\gamma+\frac{3}{2})}\exp(-E_g/K_BT)                \label{eq2}
\end{equation}

Where A is a constant, and $\gamma$ is a phenomenological exponent (arguably varying between 1/2 and 3).\cite{ref34} It is worth noting that a small variation in $\gamma$ does not affect $E_g$ at a given temperature. Combining equations (1) and (2) gives us
\begin{equation}
I = A. T^{\gamma+\frac{3}{2}}\exp[-E_g/K_BT] \left(\exp[eV/K_BT] - 1\right) \label{eq3}
\end{equation}

\begin{figure*}
\centering
\includegraphics[width=\textwidth]{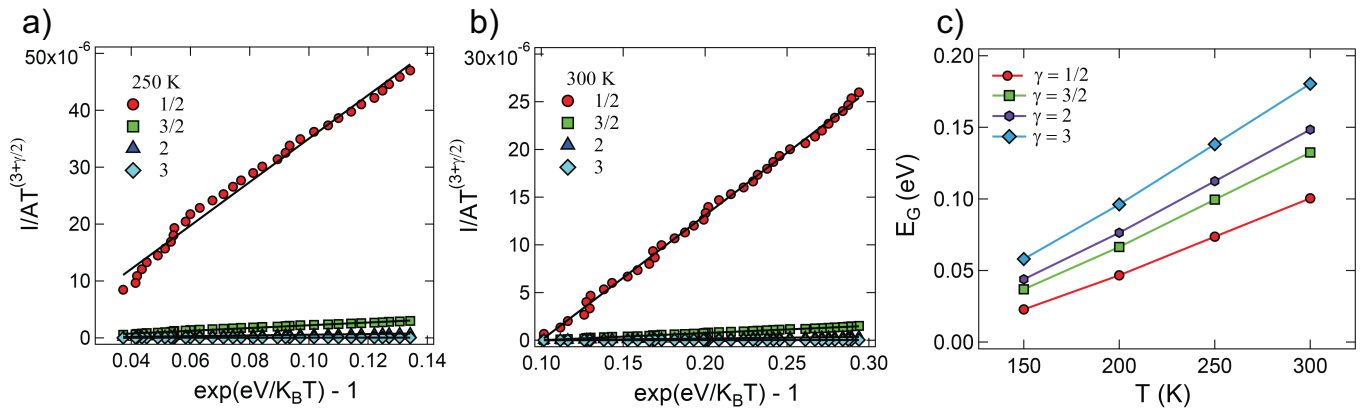} 
\caption{(color online) Estimation of gap energy $E_g$. (a-b) Plot of $I/T^{\gamma+\frac{3}{2}}$ vs. $(\exp(eV/K_BT) – 1)$ at a given temperature $T$. Equation 3 yields $E_g$ via the fitted slope of the curve for different gamma. (c) In this plot, we show the temperature dependence of gap energy $E_g$. Interestingly, $E_g$ decreases as a function of temperature. Also noticeable is a very weak dependence of $E_g$ on temperature exponent gamma.
} 
\end{figure*}

We fit experimental data using equation (3) for different values of $\gamma$. For fitting purposes, we plot  I/T$^{\gamma+\frac{3}{2}}$ as a function of $(\exp(eV/K_BT) – 1)$ at different temperatures. Basically, the equation predicts a straight line dependence of y-ordinate on the x-axis variable, the slope of which yields $E_g$, see Fig. 5a. Fig. 5b shows the resulting plot of $E_g$ vs temperature for different $\gamma$ values. Clearly, $E_g$ at a given temperature does not seem to be much affected by the variation in $\gamma$. The estimated value of $E_g$ increases modestly from $\sim$0.03 eV to $\sim$0.1 eV as temperature increases. Apparently, the gap energy exhibits opposite trend to conventional semiconductor where $\Delta$ increases as temperature decreases. Nevertheless, the gap energy is of similar magnitude as the magnetic Coulomb’s interaction between magnetic charges on honeycomb vertices in the reverse biased state,\cite{ref16} ($\mu_o/4\pi$).(3Q)$^{2}$/r $\sim$ 0.03 eV. The prevalence of high multiplicity charges (3Q) in the reverse biased state impedes electrical conduction in magnetic diode.\cite{ref16,ref30} In other words, the magnetic Coulomb’s interaction energy between 3Q charges creates a barrier for the conduction electrons.\cite{ref35} This barrier acts as the gap energy between low energy magnetic charge state (mainly comprised of $\pm$Q charges) and the energetic metastable state of $\pm$3Q charges. This new analysis further invokes the role of magnetic charge physics in magnetic diode phenomena in permalloy honeycomb lattice.

\section{Conclusion}

Finally, we summarize the findings. Magnetic diode provides one of the very few prototypes of room temperature spintronic diode, which operates at an ultra-low forward voltage $\sim$mV. The comprehensive study has not only elucidated the magnetic field and temperature tuning of magnetic diode, but has also unveiled new properties that were not known before. First of all, the unidirectional conduction, signifying the diode behavior in permalloy honeycomb lattice, is demonstrated to persist across a broad temperature range. Second, unlike a conventional diode where $V_F$ increases as a function of temperature, the forward voltage in magnetic diode actually decreases. Similarly, a contrasting behavior is also detected in the temperature dependence of gap energy $E_g$. Despite the weak semiconducting characteristic manifested by the permalloy honeycomb lattice, the surprising temperature dependencies of $V_F$ and $E_g$ suggest the onset of a new mechanism behind the unidirectional conduction in the 2D lattice. Third, a highly intriguing observation is made in the magnetic field dependence of unidirectional conduction-- we find that the diode switching carry over the remnant field effect to lower temperature. Although, the magnetic field application does not seem to affect the unidirectional conduction at higher temperature, the system preserves the remnant state when temperature is reduced, most likely comprised of low multiplicity $\pm$Q charges. Consequently, the diode behavior disappears. However, when warmed to room temperature and subsequently cooled to low temperature, the system re-establishes the unidirectional conduction. It suggests that magnetic diode preserves electrical memory. Lastly, our study has further vindicated the role of magnetic charge physics as the underlying mechanism behind the magnetic diode effect. Unlike the conventional semiconductor with band gap, the magnetic Coulomb interaction energy between magnetic charges on honeycomb vertices dictates the unidirectional electrical conduction process in permalloy honeycomb lattice. The emergent mechanism sets strong foundation for the development of next generation spintronics devices. 

\section{Experimental section}
\textit{Sample Fabrication:} Permalloy (Py) honeycomb lattice sample is nanofabricated using diblock template method, which results in large throughput sample with ultra-small connecting elements of $\sim$ 11 nm in length. Details about the synthesis of diblock template can be found somewhere else.\cite{ref31} The diblock template is fabricated by spin coating copolymer PS-b-P4VP on silicon substrate, followed by solvent vapor annealing.\cite{ref31} The resulting diblock template resembles a honeycomb lattice with a typical element size of 11 nm (length) and 4 nm (width), as is shown in Figure 1c. This topographical property was exploited to create metallic honeycomb lattice by depositing permalloy, Ni$_{0.81}$Fe$_{0.19}$, in near parallel configuration in an electron-beam evaporator. The substrate was rotated at a moderate constant speed about its axis during the deposition process to create uniformity in the film thickness.

\textit{Electrical Measurements: } Electrical measurements were performed on a 7 mm $\times$ 4 mm sample using the four-probe configuration, see inset in Fig. 1c. Electrical contacts were made using silver paste. All four contacts were in linear configuration. Electrical data at T = 300 K was also verified using a floating four-probe contact, made by hanging steel wires. Measurements at low temperature were performed on samples with silver paste contacts only, due to their robustness, in a cryogen-free 9 T magnet with a base temperature of $\sim$ 5 K. While the R vs. T measurements in zero and applied fields were performed using a highly sensitive linear resistance bridge LR720, the I-V characteristics were obtained using the synchronized set of Keithley current source meter 6242 and a Keithley nanovoltmeter 2182.

\textit{Magnetic Measurements: } Magnetic hysteresis measurements were performed in SQUID magnetometer from Quantum Design.

\medskip
\noindent\textbf{Acknowledgements} \\
DKS thankfully acknowledges the support by US Department of Energy, Office of Science, Office of Basic Energy Sciences under the grant no. DE-SC0014461.


\medskip
\noindent\textbf{Conflict of Interest} \\
The authors declare that they have no known competing financial interests or personal relationships that could have appeared to influence the work reported in this paper.

\medskip
\noindent\textbf{Data Availability Statement}\par
\noindent The data that supports the findings of the study are available from the corresponding author upon reasonable request.

\medskip
\noindent\textbf{Keywords}\par
\noindent Magnetism, Nanostructured Materials, Electrical Measurements, Magnetic diode

\medskip

%


\medskip

\end{document}